\documentstyle[twoside,fleqn,espcrc2,epsf,epsfig]{article}

\begin{document}



\begin{titlepage}

\pagenumbering{arabic}

\center{\Large{{EUROPEAN ORGANIZATION FOR PARTICLE PHYSICS}}}

\vspace*{0.5cm}

\begin{tabular}{l r}
  \hspace{10cm} CERN-OPEN/97-024 \\
  \hspace{10cm} 18 September, 1997 \\
\end{tabular}

\vspace*{2.0cm}
\center{
\Huge {\bf Interconnection Effects
in Multiparticle Production from\\
WW Events at LEP
}\\
\vspace*{0.6cm}
   {\normalsize Alessandro De Angelis}\\
\vspace*{-0.6cm}
   {\footnotesize CERN, Geneva, Switzerland} \\
\vspace*{0.5cm}
}

\center{
{\bf{Abstract}}}

\begin{center}

\end{center}

\noindent
Interactions between the products of the hadronic decays of 
different Ws in WW pair events
can occur at several stages: from the colour rearrangement between the quarks
coming from the primary branching, to the gluon exchange during the parton 
cascade, to the mixing of identical pions due to Bose-Einstein correlations. 
Besides the intrinsic interest of their study
related to the understanding of the multiparticle production 
mechanisms, these phenomena can 
affect the ultimate accuracy in the W mass measurement by LEP 
2. The status of the experimental analysis on interconnection effects between 
W pairs hadronically decaying is reviewed in this paper.
\vskip 2 truecm
\begin{center}
{\em
Presented at the XXVII International Symposium on Multiparticle Dynamics}\\
{\em Laboratori Nazionali di Frascati (Roma), September 1997.}\\
{\em To be published in the Proceedings.}

\end{center}

\end{titlepage}

\setcounter{page}{1}    
\def\asmz{$\alpha_s(M_Z)$}
\def\ass{\alpha_s(\sqrt{s})}
\newcommand{\kos}{\ifmmode {{\mathrm K}^{0}_{S}} \else
${\mathrm K}^{0}_{S}$\fi}
\newcommand{\kpm}{\ifmmode {{\mathrm K}^{\pm}} \else
${\mathrm K}^{\pm}$\fi}
\newcommand{\ko}{\ifmmode {{\mathrm K}^{0}} \else
${\mathrm K}^{0}$\fi}
\def\as{$\alpha_s$}
\def\asb{$\alpha_s\sp{b}$}
\def\asc{$\alpha_s\sp{c}$}
\def\asuds{$\alpha_s\sp{udsc}$}
\def\Lam{$\Lambda$}
\def\ZP{Z.\ Phys.\ {\bf C}}
\def\PL{Phys.\ Lett.\ {\bf B}}
\def\PR{Phys.\ Rev.\ {\bf D}}
\def\PRL{Phys.\ Rev.\ Lett.\ }
\def\NP{Nucl.\ Phys.\ {\bf B}}
\def\CPC{Comp.\ Phys.\ Comm.\ }
\def\NIM{Nucl.\ Instr.\ Meth.\ }
\def\Coll{Coll.,\ }
\def\Rmu{$R_3(\mu)/R_3(had)$\ }
\def\Re{$R_3(e)/R_3(had)$\ }
\def\Rmue{$R_3(\mu +e)/R_3(had)$\ }
\def\ee{$e\sp{+}e\sp{-}$}
\newcommand{\Wfj}{WW $\rightarrow 4jets$ events}
\newcommand{\Wtj}{WW $\rightarrow 2jets \, \ell \bar{\nu}$ events}


\title{Interconnections Effects 
in Multiparticle Production from\\ 
WW Events at LEP}
\author{Alessandro de Angelis
\address{CERN, Geneva, Switzerland}}

\begin{abstract}
Interactions between the products of the hadronic decays of 
different Ws in WW pair events
can occur at several stages: from the colour rearrangement between the quarks
coming from the primary branching, to the gluon exchange during the parton 
cascade, to the mixing of identical pions due to Bose-Einstein correlations. 
Besides the intrinsic interest of their study
related to the understanding of the multiparticle production 
mechanisms, these phenomena can 
affect the ultimate accuracy in the W mass measurement by LEP 
2. The status of the experimental analysis on interconnection effects between 
W pairs hadronically decaying is reviewed in this paper.
\end{abstract}

\maketitle


\section{Introduction}

The data collected at LEP 
are at the highest available centre-of-mass energies in $e^+e^-$ interactions,
and a new physics window is opened: the production and decay of W boson
pairs. 

The possible presence of interference (due to colour reconnection
and Bose-Einstein correlations (see for 
example~\cite{sjre,were,elre,sjbe,moller} and
~\cite{yellow} for a review)
in hadronic decays of WW pairs has been discussed on a theoretical basis, 
in the framework of the measurement of the W mass: 
this interference~\cite{yellow} can induce a systematic uncertainty on the W 
mass measurement in the 4-jet mode which is of the order of 40 MeV, i.e., 
comparable with the expected accuracy of the measurement. 

Interconnection can happen due to the fact that the typical lifetime of the W
($\tau_W \simeq \hbar /\Gamma_{W} \simeq 0.1$ fm/$c$) is one order of magnitude 
smaller than the typical hadronization times.
The interconnection between the products of the hadronic decays of 
different Ws in WW pair events
can occur at several stages: 
\begin{enumerate}
\item from colour rearrangement between the quarks
coming from the primary branching, 
\item due to gluon exchange during the parton 
cascade, 
\item in the mixing of identical pions due to Bose-Einstein correlations.
\end{enumerate}

The first two enter in the category of the QCD effects. The QCD interference 
effects can mix up the two colour singlets and produce hadrons that cannot be 
uniquely assigned to either W. The perturbative effects are suppressed by the 
need to exchange two gluons and conserve the colour; this creates a suppression 
($1/N^2_C - 1$), such that the effect is expected to be small (and to induce a 
possible shift of only about 5 MeV in the W mass). Nonperturbative effects are more 
difficult to compute, and they need models; typically, 
according to models, 
such effects can induce shifts of the order of 
40 MeV in the W mass~\cite{yellow,nova}.

The study of Bose-Einstein Correlations (BEC) is complicated by the fact that a 
complete description would need the symmetrization of the amplitude for the 
multiparticle system, which is computationally 
difficult. One must thus make approximations 
and build models~\cite{fialko}; care should be taken not to 
distort the multiplicity distribution by increasing the probability of final 
states with large multiplicity (the bounds form the precise $R_b$ measurements 
are in this sense very stringent). The subject became very popular due to the 
prediction~\cite{sjbe} that BEC could shift by 100 MeV the measured W mass.

How to investigate experimentally interconnection effects ? 
The WW events allow a comparison of
the characteristics of the W hadronic decays when
both Ws decay in fully hadronic modes 
(in the following I shall often 
refer to this as to the $(4q)$ mode)
with the case in which the other W decays
semileptonically
($(2q)$ mode for brevity). These 
should be equal in the absence of interference between the
products of the hadronic decay of the Ws.
\begin{itemize}
\item A qualitative argument shows that
the effect of colour reconnection
between the decay products of
different Ws could affect the charge multiplicity:
close to the WW threshold, the presence of two cross-talking
dijets in the fully hadronic WW decay allows the evolving
particle system to have a larger kinetic energy than in the
case of independent dijets with no cross-talk.
In a recently proposed model~\cite{elre}, this
correlation could lead to a 
charge multiplicity for $(4q)$ events that is significantly smaller than 
twice the multiplicity of $(2q)$ events.
In addition, effects of colour reconnection are expected to be considerably
enhanced in kinematic regions where there is strong overlap
between jets originating from different Ws.

One could be more sensitive to interconnection effects by studying inclusive 
particle distribution ``oriented'' by the event axis (like thrust or rapidity). 
However, a severe experimental warning is given by the fact that WW events are 
selected mostly by topological cuts, which could create possible biases between 
the $(4q)$ and the $(2q)$ in the definition of the axis.
\item For what is related to BEC, 
the queen of the observables is the two-particle correlation function.
BEC could anyway slightly increase the multiplicity for $(4q)$ events in
some models~\cite{moller}.
\end{itemize}

\section{Experimental Results}
The cross section for WW production at 
$\sqrt{s} = 161$ GeV being small (about 3 pb) and the
data taking at 183 GeV having just started,
the experimental results are based on the analysis of the data taken at 
172 GeV. 
For this energy the WW cross section is about 12 pb; the main background is 
given by $q\bar{q}$
events, with a cross section (for an effective center of mass 
energy larger than 10\% of the maximum annihilation energy) of about 125 pb.
The situation will improve with the analysis of the data 
at 183 GeV. A total integrated luminosity of 50 pb$^{-1}$ per experiment can be 
expected before end 1997.
The WW cross section is about 15 pb; the cross section for
$q\bar{q}$ events is about 105 pb.
\begin{itemize}
\item
About 4/9 of the WW are \Wfj. At threshold, 
their topology is a clear 4-jet back-to-back topology, with no 
missing energy; the constrained invariant mass of two
jet-jet systems equals the W mass. Still at 172 and 
183 GeV these 
characteristics allow a clean selection, such that the typical 
efficiency is about 85\%, for a purity of about 80\%.
\item
About 4/9 of the WW are \Wtj. At threshold, 
their topology is a clear 2-jet back-to-back topology, with a lepton and
missing energy opposite to it; 
the constrained invariant mass of the
jet-jet system and of the lepton-missing energy system 
equals the W mass. Still at 172 and 
183 GeV, despite of the boost, these 
characteristics allow a clean selection, such that the typical 
efficiency is about 80\%, for a purity of about 90\%.
\end{itemize}
For a more detailed description of the selection criteria and of the cross 
section and branching fractions measurements, see~\cite{hartmann}.

\subsection{QCD Effects}

Charged multiplicity and momentum spectrum of charged particles in 
\Wfj~and \Wtj~have been studied by DELPHI~\cite{delmul,delpre}, OPAL
~\cite{opamul} and L3~\cite{l3mul} looking for possible effects of correlations 
on the \Wfj.

In a recent OPAL paper~\cite{opamul}, inclusive distributions of charged 
particles from the decay of the W in WW events are studied. The main 
conclusions are that:
\begin{itemize}
\item The momentum spectrum of charged particles in \Wfj~and in \Wtj~ agree with 
the expectations from PYTHIA~\cite{lund}, 
KORALW~\cite{KORALW} and HERWIG~\cite{HERWIG} (figure \ref{xpo}). 
\begin{figure*}
\begin{center}
\mbox{\epsfxsize11.8cm\epsffile{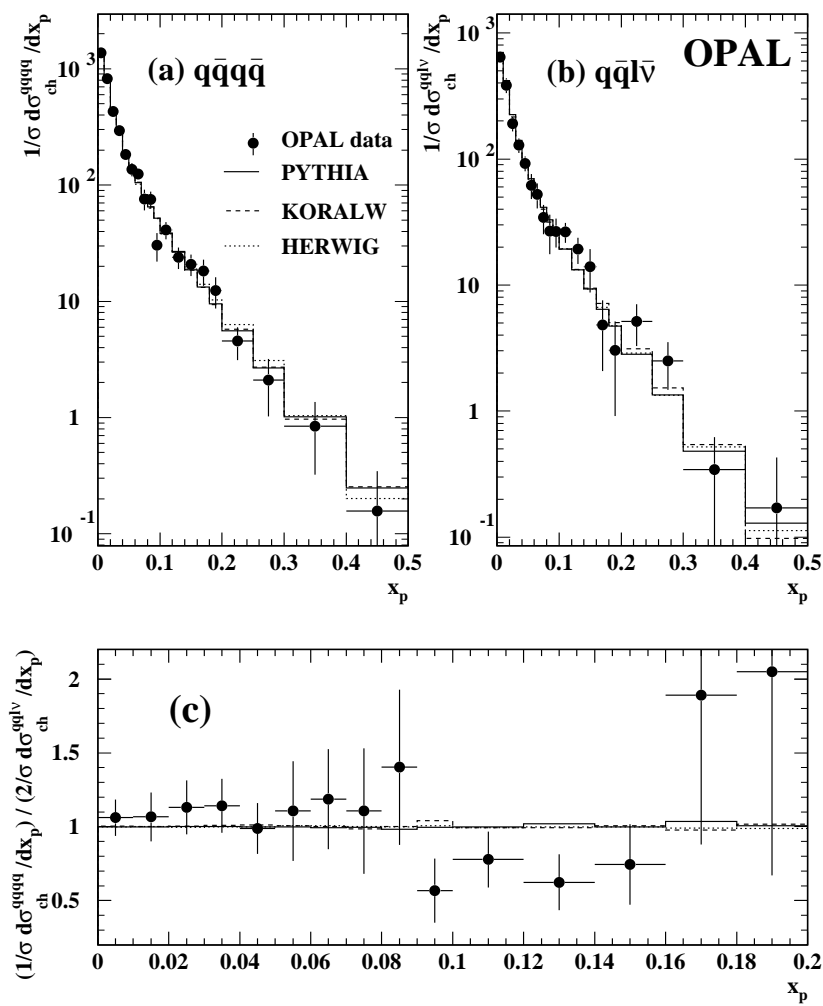}}
\end{center}
\caption[]{\label{xpo}
Corrected $x_p$ distributions for (a) $(4q)$ events and (b) $(2q)$ events.
The ratio between $(4q)$ and twice $(2q)$ is shown in (c).}
\end{figure*}
No interconnection effects were 
present in the \Wfj~ simulated using these Monte Carlo programs.
\item The mean values of two oriented event shape variables:
\begin{eqnarray*}
 <1-T>^{(4q)} & = & 0.227 \pm 0.036 \pm 0.021\\
 <|y|>^{(4q)} & = & 1.033 \pm 0.042 \pm 0.025
\end{eqnarray*}
in $(4q)$~are also in agreement with the above models without interconnection 
effects.
\item The average momentum fraction $<x_p>$ 
(where $x_p = 2p/\sqrt{s}$) for charged particles in 
$(4q)$~ and in $(2q)$~are also consistent with models, and within 1.5$\sigma$ between 
each other:
\begin{eqnarray*}
 <x_p>^{(4q)} & = & (3.22 \pm 0.13 \pm 0.08) \times 10^{-2}\\
 <x_p>^{(2q)} & = & (3.60 \pm 0.20 \pm 0.11) \times 10^{-2}\\
 <\Delta x_p> & = & (-0.38 \pm 0.25) \times 10^{-2} \, ,
\end{eqnarray*}
\end{itemize}
where $\Delta x_p = x_p^{(4q)}-x_p^{(2q)}$. 

\begin{figure}
\begin{center}
\mbox{\epsfxsize6.0cm\epsffile{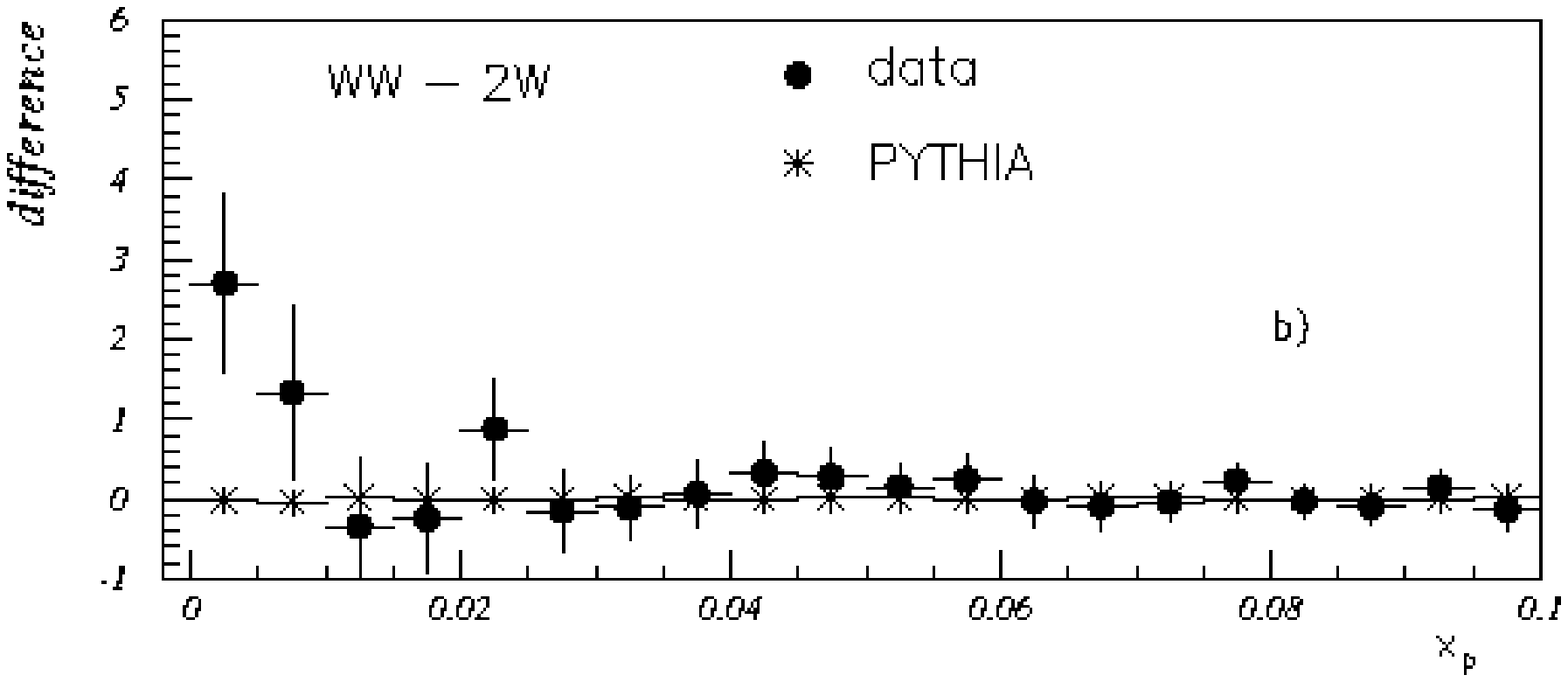}}
\mbox{\epsfxsize6.0cm\epsffile{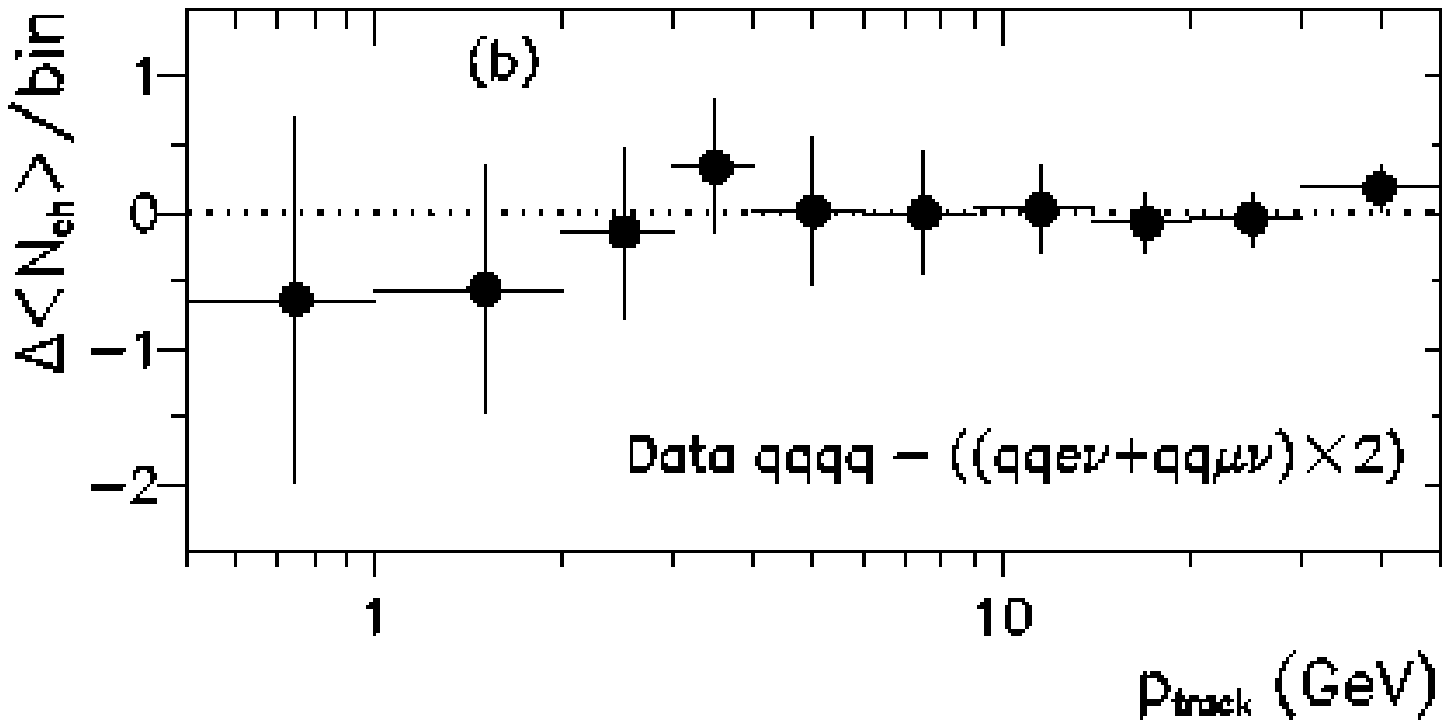}}
\end{center}
\caption[]{\label{xp} Difference between the 
$x_p$ distribution for $(4q)$ and twice $(2q)$ as 
measured by DELPHI (top) and L3 (bottom).}
\end{figure}

One could try to increase the sensitivity to interconnection effects by 
subtracting from the charged momentum 
distribution in \Wfj~twice the momentum spectrum in \Wtj. In a preliminary 
paper~\cite{delpre}, DELPHI observes a $\sim 2\sigma$ effect in the low-$x_p$
region; there is an excess of 
of $3.9 \pm 1.5 \pm 0.7$ charged particles for 
$x_p < 0.01$ (figure \ref{xp}a). This result is not confirmed by L3~\cite{l3mul}, 
which 
observes an excess in the opposite direction (figure \ref{xp}b)
nor by OPAL~\cite{opamul} (figure \ref{xpo}c).

\begin{table}
\begin{center}
\small{
\begin{tabular}{|l|l|l|}
\hline
                     &    $(4q)$   &  $(2q)$ \\
\hline
  {\em L3 preliminary~\cite{l3mul}}  & {\em 37.9$\pm$0.9} & {\em 19.4$\pm$0.6}  \\
\hline\hline
  DELPHI~\cite{delmul} &     40.6$\pm$1.7$\pm$1.1    &  18.0$\pm$0.9$\pm$0.6  \\
\hline
  OPAL~\cite{opamul}  & 38.3$\pm$1.1$\pm$0.6     &  18.4$\pm$0.9$\pm$0.4  \\
\hline\hline
  Average &   38.94$\pm$1.07   &     18.22$\pm$0.73\\
\hline
\end{tabular}}
\end{center}
\caption{\label{tabmul}Mean charged 
multiplicities in the $(4q)$ and in the $(2q)$ 
channels. The L3 value was not used in the calculation of the average, since
no systematic error was specified.}
\end{table}
Finally, the measurements of charge 
multiplicity in $(4q)$~and $(2q)$~are consistent
(table \ref{tabmul}). One obtains
\begin{equation}
 <n>^{(4q)} - 2<n>^{(2q)} = 2.5 \pm 1.8 (stat + syst) \, .
\end{equation}

In conclusion, the studies of mean charged multiplicity and inclusive 
particle distribution do not indicate at the present level of statistics
the presence of interconnection effects. 

\begin{figure}
\vspace*{0.5cm}
\begin{center}
\mbox{\epsfxsize6.0cm\epsffile{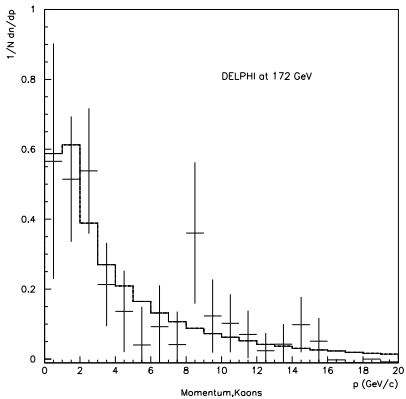}}
\end{center}
\vspace*{-0.5cm}
\caption[]{\label{kid} Momentum spectrum of identified K$^{\pm}$ in
$(4q)$ events, compared to the expectation from PYTHIA (solid line).}
\end{figure}

A study which can be very fruitful in the future is the comparison of
the momentum spectra for identified hadrons (heavier than pions), for which the 
multiplicity suppression at low $x$ is expected to be stronger than for 
the inclusive (pion-dominated) spectrum~\cite{val}. A preliminary analysis
by DELPHI~\cite{yasser}, 
based on the 1996 data, shows agreement with PYTHIA without 
interconnection effects (figure \ref{kid}); the
errors are completely dominated by statistics.

\subsection{Bose-Einstein Correlations}
BEC manifest themselves in an enhancement in the 
production of pairs of identical bosons close in phase space.

To study the enhanced probability for emission of two identical bosons,
the correlation function $R$ is used as a probe. For pairs of particles, 
it is defined as 
\begin{equation}
R(p_{1},p_{2}) = \frac{P(p_{1},p_{2})}{P_{0}(p_{1},p_{2})} \, ,
\end{equation}
where $P(p_{1},p_{2})$ is the two-particle probability density, subject to
Bose-Einstein symmetrization, $p_{i}$ is the four-momentum of
particle $i$, and $P_{0}(p_{1},p_{2})$ is a
reference two-particle distribution which, 
ideally, resembles $P(p_{1},p_{2})$ in all respects, apart from the lack
of Bose-Einstein symmetrization. 

If $f(x)$ is the space-time distribution of the source, $R(p_{1},p_{2})$ 
takes the form
\begin{eqnarray*}
R(p_{1},p_{2})=1+|G[f(x)]|^{2} \, ,
\end{eqnarray*}
where $G[f(x)]=\int{f(x)e^{-\imath(p_{1}-p_{2})\cdot x \,} dx}$ 
is the Fourier transform of $f(x)$. 
Thus, by studying the correlations between the momenta of pion pairs,
one can determine the distribution of the points of origin of the pions.
Experimentally, 
the effect is often described in 
terms of the variable $Q$, defined by 
$Q^{2}=M^{2}(\pi \pi)-4m^{2}_{\pi}$,
where $M$ is the invariant mass of the two pions.
The correlation function can then be written as
\begin{equation}
R(Q) = \frac{P(Q)}{P_{0}(Q)},
\end{equation}
which is frequently parametrized by the function  
\begin{equation}\label{beq}
R(Q)= 1 + \lambda e^{-r^2Q^2} \, .
\end{equation}
In the above equation, in the hypothesis of a pion source spherically symmetric,
the parameter $r$ gives the RMS radius source, and $\lambda$ is
the strength of the correlation between the pions. DELPHI~\cite{delbe}
has shown that by taking only primary pions from $e^+e^-$ 
annihilation $\lambda$ 
is consistent with 1, and thus $\lambda$ 
can be interpreted as the fraction of interacting pairs.
The data from $e^+e^-$ annihilations from PEP energies to LEP show values of $r$ 
around 0.5 fm; the value of $\lambda$ strongly depends on the analysis 
technique, ranging from 0.2 to 1.
\begin{figure*}
\begin{center}
\mbox{\epsfxsize12.0cm\epsffile{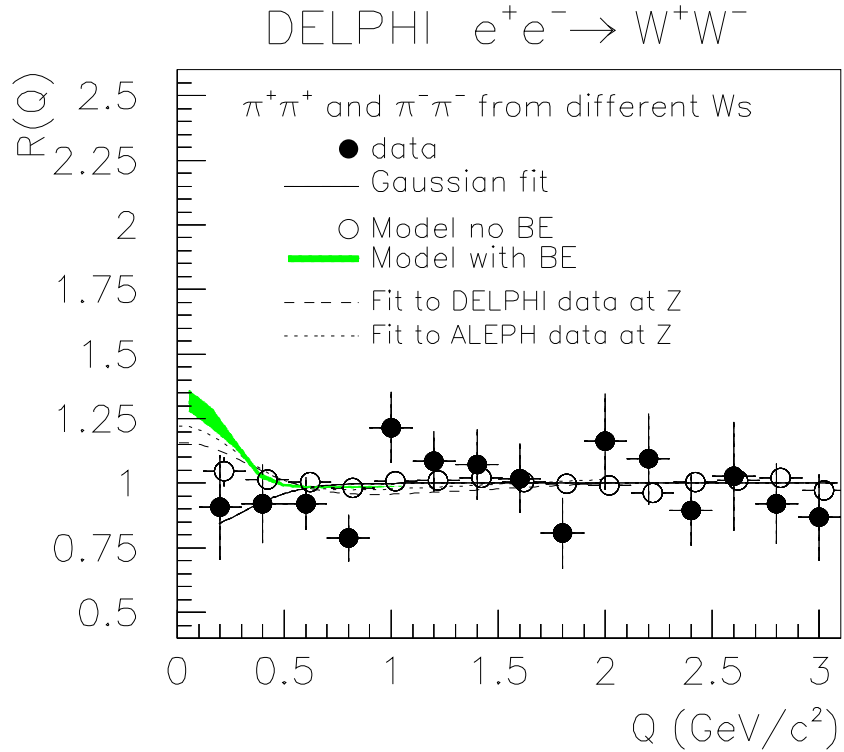}}
\end{center}
\caption[]{\label{amir} 
The correlation function $R(Q)$ for like-sign particles arising from
different Ws for data ({closed circles}) and simulated events 
without Bose-Einstein symmetrization ({open circles}).
The {shaded area} represents the model prediction for events with
Bose-Einstein  symmetrization (see text).
The {solid curve} shows the result of the fit using equation (\ref{beq}).
 The {dashed and dotted curves} are
results of fits to $R(Q)$ distributions for like-sign particles 
measured in Z decays.}
\end{figure*}

It can be understood from what said above that in the study of BEC the main 
problem is given by a good choice of the reference sample. Normally three 
reference samples are used in the literature:
\begin{itemize}
\item Pairs of particles of opposite sign. The drawback of this choice is that 
many correlated unlike sign pairs come from resonance decays, and the influence 
of those on the correlation function has to be corrected for by means of a
simulation (with possible biases).
\item Pairs of particles taken from artificial events constructed by mixing 
particles from different real events. The mixing technique introduces 
arbitrarity, and possible biases.
\item Monte Carlo events simulated without BEC. 
\end{itemize}

Four analyses are proposed for detecting BEC between 
like-sign pions from different Ws.
DELPHI presented two analyses on the subject, one published~\cite{amiran} 
and one
preliminary~\cite{delpre}. ALEPH presented also two analyses~\cite{alepre}.
All are based on the sample of 10 pb$^{-1}$ collected
at $\sqrt{s} = 172$ GeV.
\begin{enumerate}
\item DELPHI~\cite{amiran} proposes a technique which uses the unlike-sign 
reference sample without need of the simulation for 
correcting for the effect of resonances (provided the colour reconnection 
effects are not large).

To obtain the two-particle $Q$ distribution for pairs of pions coming 
from different Ws, the following procedure can be
used. The $Q$ distribution for pion pairs is measured in
$(4q)$ events.
This distribution is the sum of the
distribution
of pion pairs coming from the same W
and of that of pion pairs coming from different Ws. 
The contribution of pairs coming from the same W is subtracted
statistically, using the
$Q$ distribution obtained from $(2q)$ events.
The same procedure is followed for both like-sign
and unlike-sign pion pairs to obtain $P(Q)$ and $P_{0}(Q)$, respectively. 

In the absence of colour reconnection,
the two-particle density for unlike-sign pairs
should be identical to the like-sign distribution
except for BEC effects, and should not
contain pairs from decays of particles or resonances.
The ratio of like-sign to unlike-sign pairs: 
\begin{equation}\label{delbel}
 R(Q) = \frac{P^{(4q)}_{++}(Q) - 2P^{(2q)}_{++}(Q)}
  {P^{(4q)}_{+-}(Q) - 2P^{(2q)}_{+-}(Q)}
\end{equation} 
is shown in 
figure \ref{amir}.

\begin{figure}
\vspace*{0.5cm}
\begin{center}
\mbox{\epsfxsize7.0cm\epsffile{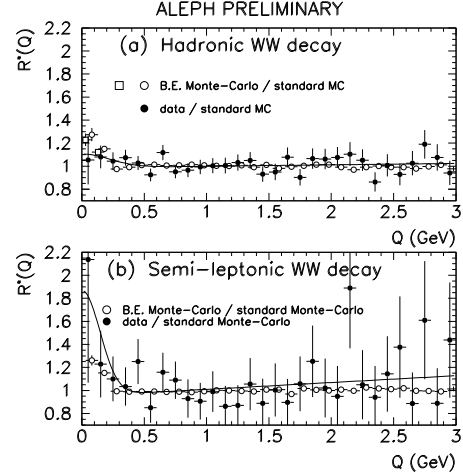}}
\end{center}
\vspace*{-2.5cm}
\caption[]{\label{aleph1}
Ratio of both data (full circles) and Monte Carlo with BEC (open circles)
to the Monte Carlo without BEC for the hadronic and semi-leptonic W decays. The
open squares in (a) show the Monte Carlo with BEC in each W decay only. The 
solid line
shows the result of the fit to the data.}
\end{figure}
\begin{figure}
\vspace*{0.5cm}
\begin{center}
\mbox{\epsfxsize7.0cm\epsffile{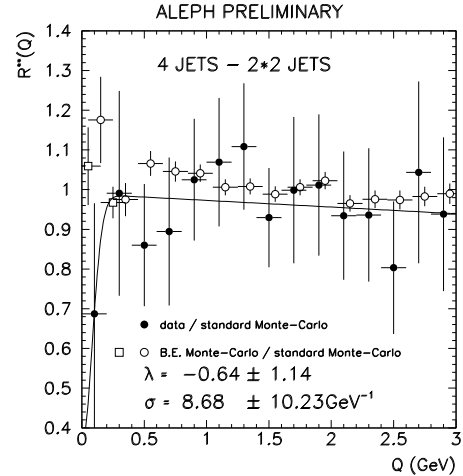}}
\end{center}
\vspace*{-2cm}
\caption[]{\label{aleph2}
The ratio of the data (full circles) and of the Monte Carlo with BEC (open 
circles) to the Monte Carlo without BEC for pion pairs from different W.
Open squares as in the previous figure.}
\end{figure}

Since no enhancement of the correlation function is observed at low
$Q$ values, the para\-meter $r$ is not well defined. 
The fit to the correlation function with expression (\ref{beq}) 
was therefore
performed with a fixed value of 
$r$=0.5 fm, as measured in Z decays.
The fit (shown by the solid line) yielded the value:
\begin{eqnarray}
\lambda=-0.20 \pm 0.22 (stat) \pm 0.08 (syst) \, .
\end{eqnarray}
At the present level of statistics, no evidence is observed for Bose-Einstein 
correlations between pions from {different} Ws. If the effect were 
present, it would be visible (shaded area in the figure); data do not show it 
at $2\sigma$.
Figure \ref{amir} also shows $R(Q)$ distributions predicted 
using WW events generated by PYTHIA with the LUBOEI routine~\cite{lund}.
When no Bose-Einstein effects are included, the correlation function $R(Q)$ 
is found to be equal to one 
within errors in the whole $Q$-region presented,
in good agreement with the data (open circles). 
\item Another DELPHI analysis~\cite{delpre} studies the correlation function 
between same-sign pions in $(4q)$ and in $(2q)$ events, using a reference sample 
from event mixing. If the pions from different Ws were completely correlated, 
$R(Q)$ would be the same in $(2q)$ and in $(4q)$; otherwise, it would be larger 
in $(2q)$ (since the pars of uncorrelated pions in $(4q)$ would dilute the 
effect).

Within the limited statistics, the data disfavor the complete correlation 
scenario.
\item A first ALEPH analysis~\cite{alepre} starts from the tuning of the 
simulation to reproduce the two-particle correlation in the data (about 
1 pb$^{-1}$) taken at the Z in 1996 for calibration purposes. This is done by 
introducing in the simulation BEC according to the technique described in 
~\cite{jada}.
Three two-particle correlation functions $P(Q)$
can then be constructed from Monte Carlo in $(4q)$:
\begin{itemize}
\item A sample $P_0(Q)$ without BEC;
\item A sample $P^{(1)}(Q)$ with BEC between all pion pairs;
\item A sample $P^{(1/2)}(Q)$ with BEC only between $\pi$ from the same W.
\end{itemize}
The last two coincide in $(2q)$ events.
The experimental results are shown in figure \ref{aleph1}; the
function $P^{(1/2)}(Q)/P_0(Q)$ (open squares) provides the best 
approximation of the experimental result $P(Q)/P_0(Q)$ (closed circles).
The data favour the hypothesis of no cross-talk between pions from different Ws.
\item Finally, a second ALEPH analysis~\cite{alepre}
is similar to the DELPHI analysis described at item 1. 
A correlation function defined as in Eq. \ref{delbel} is computed, and then 
normalized to the same function from the simulation without BEC to remove 
possible residual 
biases. The results are shown in figure \ref{aleph2}; again, the data 
disfavor the hypothesis of BEC between $\pi$ from different Ws.
\end{enumerate}

The general conclusion on BEC is that two experiments 
disfavor (one at the $2\sigma$ level, the other at the $1\sigma$ level) the 
hypothesis of complete BEC between pions from different Ws.
\begin{figure*}
\centerline{\epsfig{file=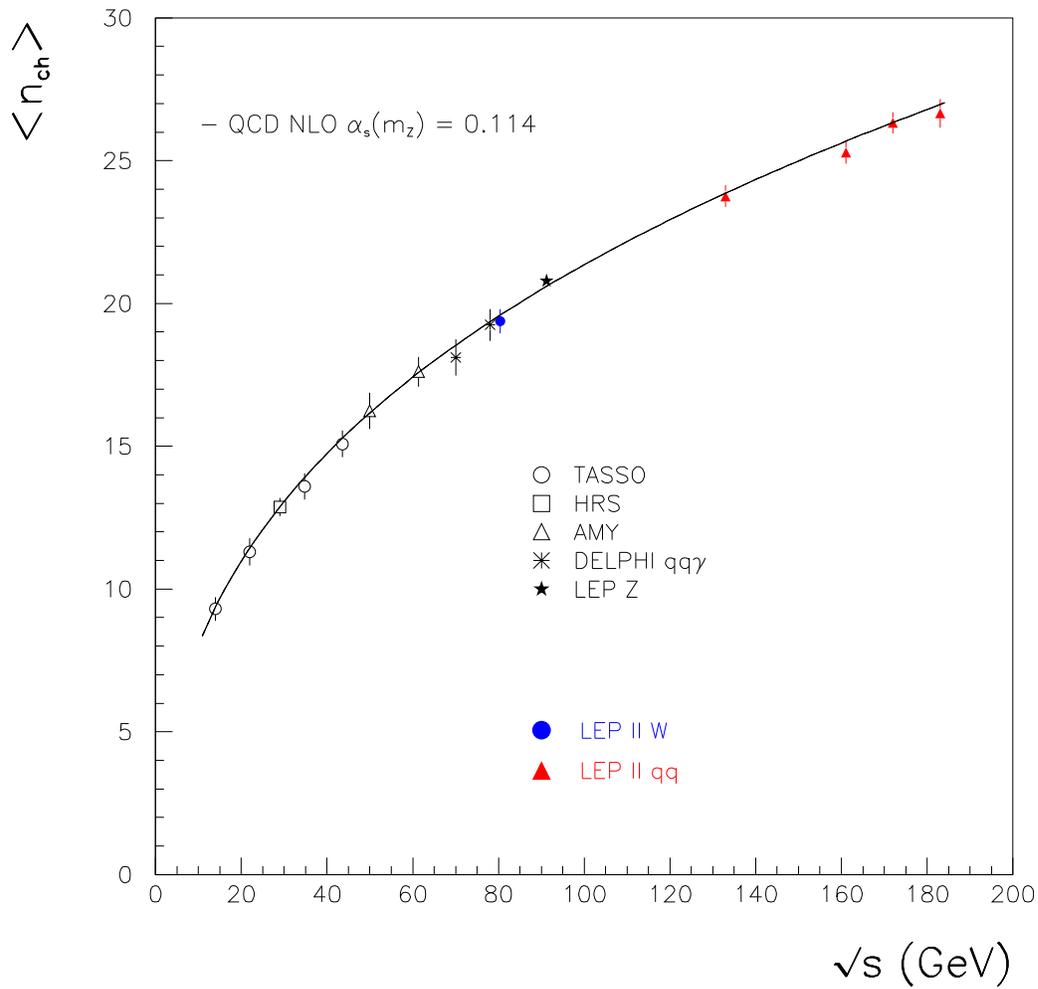,width=0.95\linewidth}}
\caption[]{Measured charged multiplicity
in $e^+e^- \rightarrow {\mathrm q}\bar{\mathrm q}$ events as a function
of centre-of-mass energy $\sqrt{s}$. The LEP II results on the W
are compared with other experimental results
and with a fit to a prediction from
QCD in Next to Leading Order.
The measurements have been corrected for the different
proportions of $b\bar{b}$ and $c\bar{c}$ events at the various energies.}
\label{mulea}
\end{figure*}

\section{Conclusions}

At the present level of statistics there is no evidence for Bose-Einstein
correlations between pions originating from different Ws in
\Wfj. A complete Bose-Einstein correlations between such pions is 
unlikely.

From the results obtained on hadronic W decays, we observe that the charged
multiplicity from WW systems in which both Ws decay hadronically
is consistent with twice that from a
W  whose partner decays semileptonically:
\begin{equation}
\frac{2<n>^{(2q)}}{<n>^{(4q)}} =
0.936 \pm 0.045 \, .
\end{equation}
As discussed in~\cite{delmul,opamul},
this is one of the facts ruling out the model~\cite{elre}, 
which predicts a 10\% excess in multiplicity for the $(2q)$
channel. However, at this level of statistics only such extreme 
colour reconnection models can be probed.

The studies of charged multiplicity and inclusive 
particle distributions do not indicate at the present level of statistics
the presence of interconnection effects. One can average the $(4q)$~and the 
$(2q)$~events to obtain the best value of the mean charged multiplicity in W 
hadronic decays:
\begin{equation}
 <n_W> = 19.03 \pm 0.43 (stat + syst) \, .
\end{equation}
The value of $<n_W>$ is also plotted in Figure
\ref{mulea} at an energy value corresponding to the W mass, together with a
recent compilation of $e^+e^-$ data at different
centre-of-mass energies~\cite{abreu};
it has been increased by 0.35 units to account
for the different proportion of $b\bar{b}$ and $c\bar{c}$ events 
in W decays than in continuum $e^+e^-$ events~\cite{dea}.
It lies on the same curve as the $e^+e^-$ data, consistent with the prediction 
from QCD including NLO corrections~\cite{weber}.

It should be underlined that analyses based on a WW statistics 
larger by one order of magnitude will be possible in one year from now, and 
important news, which could affect our understanding of multiparticle production
mechanisms and of the ultimate accuracy on the W mass measurement, can be
expected very soon.

\subsection*{Acknowledgements}

I thank the organizing committee of the Symposium for the nice and 
fruitful atmosphere in Frascati.
I could understand a bit more on the subject thanks to discussions with 
K. Fialkovski, A. Giovannini, V. Khoze
and P. Nason. Part of the material used for preparing this talk and this 
report comes from
P. Abreu, C. Hartmann, D. Treille and L. Vitale, to which I am grateful.

\vskip 2 truecm

\def\Abreu{DELPHI Coll., P. Abreu {\it et al.,}\ }
\def\etal{{\it et al.,}\ }

\end{document}